\documentclass[showpacs,aps,graphicx]{revtex4}
\usepackage{amsmath}
\usepackage{mathrsfs}
\usepackage{amsfonts}
\usepackage{amssymb}
\usepackage{graphicx}
\usepackage{caption}
\usepackage{eufrak}
\usepackage{multirow}
\usepackage{float}
\usepackage[colorlinks,linkcolor=blue,anchorcolor=blue,citecolor=blue]{hyperref}
\usepackage{soul,color,xcolor}
\soulregister\upcite7
\soulregister\citep7
\soulregister\citet7
\soulregister\ref7
\soulregister\pageref7
\soulregister\and7
\soulregister\Name7

\begin{document}

\title{The simplified quantum circuits for implementing quantum teleportation}

\author{Wen-Xiu Zhang\textsuperscript{1}, Guo-Zhu Song\textsuperscript{2}, and  Hai-Rui Wei\textsuperscript{1}\footnote{Corresponding author: hrwei@ustb.edu.cn} }

\address{1 School of Mathematics and Physics, University of Science and Technology Beijing, Beijing 100083, China \\
%
%
2 College of Physics and Materials Science, Tianjin Normal University, Tianjin 300387, China}

\date{\today }

\begin{abstract}

It is crucial to design quantum circuits as small as possible and as shallow as possible for quantum information processing tasks. We design quantum circuits with simplified gate-count, cost, and depth for implementing quantum teleportation among various entangled channels. Here the gate-count/cost/depth of the Greenberger-Horne-Zeilinger-based quantum teleportation is reduced from 10/6/8 to 9/4/6, the two-qubit-cluster-based quantum teleportation is reduced from 9/4/5 to 6/3/5, the three-qubit-cluster-based quantum teleportation is reduced from 12/6/7 to 8/4/5, the Brown-based quantum teleportation is reduced from 25/15/17 to 18/8/7, the Borras-based quantum teleportation is reduced from 36/25/20 to 15/8/11, and the entanglement-swapping-based quantum teleportation is reduced from 13/8/8 to 10/5/5. Note that, no feed-forward recover operation is required in the simplified schemes. Moreover, the experimentally demonstrations on IBM quantum computer indicate that our simplified and compressed schemes can be realized with good fidelity.

Keywords: Quantum teleportation, Quantum circuit compression, Quantum circuit, Quantum entanglement

\end{abstract}

\pacs{03.67.Lx, 03.65.Ud, 03.67.Mn}

\maketitle

\newcommand{\upcite}[1]{\textsuperscript{\textsuperscript{\cite{#1}}}}

\section{Introduction}\label{sec1}

Quantum information science and technologies\upcite{book,1,1-1,1-2} could outperform their classical counterparts on solving some certain tasks, and it mainly consists of quantum communication\upcite{2,2-1,2-2,2-3,2-4} and quantum computing.\upcite{3,3-1,3-2,3-3} Quantum communication is a promising and flourishing research for its security and efficiency. Many theoretical and experimental works have been proposed on quantum key distribution,\upcite{quantum-key-distribution1,quantum-key-distribution2,quantum-key-distribution3} quantum teleportation,\upcite{6,7,8} quantum secure direct communication,\upcite{secrue-direct-communication1,secrue-direct-communication2,secrue-direct-communication3} entanglement purification and concentration,\upcite{entanglement-purification-and-concentration1,entanglement-purification-and-concentration2,entanglement-purification-and-concentration3,entanglement-purification-and-concentration4,entanglement-purification-and-concentration5,entanglement-purification-and-concentration6} quantum dense coding\upcite{quantum-dense-coding1,quantum-dense-coding2,quantum-dense-coding3} and quantum secret sharing.\upcite{quantum-secret-sharing1,quantum-secret-sharing2,quantum-secret-sharing3} Quantum teleportation has proved to be a powerful tool in quantum communication, it is essential to measurement-device-independent quantum key distribution,\upcite{quantum-key-distribution} device-independent quantum secure direct communication\upcite{quantum-secure-direct-communication1,quantum-secure-direct-communication2,quantum-secure-direct-communication3} and quantum repeater.\upcite{quantum-repeater1,quantum-repeater2}

Quantum teleportation was first proposed by Bennett et al.\upcite{6} to transmit unknown single-qubit quantum states from one location to another among Einstein-Podolsky-Rosen (EPR) channels, and was late experimentally demonstrated by Bouwmeester et al..\upcite{bell} In pursuit of higher communication efficiency, quantum teleportation via different channels have been widely studied.  Compared with the traditional scheme, the teleportation via Greenberger-Horne-Zeilinger (GHZ) state channels\upcite{GHZ,partical-entangled-GHZ2} is more secure. The teleportation via cluster state channels\upcite{w1,cluster-state} makes the teleportation harder to be destroyed by single-bit measurement or local measurements risking. For a high degree of entanglement, Brown state and Borras state were used as quantum channel.\upcite{old} Subsequently, these were extended to teleport a multi-qubit state\upcite{w3,cluster-state2,cluster-state3,bell-channels} and the controlled quantum teleportation.\upcite{controlled-bidirectional,controlled-n-qubit-teleportation}  Bidirectional quantum teleportation,\upcite{Bidirectional,Bidirectional-teleportation1,Bidirectional-teleportation2} high-dimensional quantum teleportation\upcite{high-dimensions,Experimental-High-Dimensional} and experimental quantum teleportation\upcite{1Experimental-satellite,3Experimental-microwaves,quantum-teleportation1} were experimentally demonstrated recently.

Quantum entangling gates,\upcite{gate1,gate2,gate3,gate4} such as controlled-NOT (CNOT) gate or controlled-Z gate, play a prominent role in quantum computation and quantum communication. For example,
two Hadamard gates, two CNOT gates, one control-X gate and one control-Z gate are sufficient to realize the teleportation.\upcite{circuit} Nine Hadamard gates, five CNOT gates, and five control-Z gates are required to teleport arbitrary two-qubit via four-qubit cluster.\upcite{experiment-cluster1}  Five Hadamard gates, twelve CNOT gates and four control-Z gates are necessary to teleport arbitrary two-qubit via five-qubit cluster.\upcite{quantum-teleportation1} We note that lowering gate-count, cost and depth in the quantum circuit is better from the perspective of noise resiliency and resource-saving. Thus, designing quantum circuits with optimized gate-count, cost, and depth is more conducive to the realization in quantum information processing.

In this paper, we first simplify quantum circuits for teleporting arbitrary single-qubit message via GHZ state, two-qubit cluster state, three-qubit cluster state, Brown state, Borras state, and entanglement swapping, respectively. The gate-count/cost/depth of the circuits is reduced from 10/6/8 to 9/4/6, from 9/4/5
to 6/3/5, from 12/6/7 to 8/4/5, from 25/15/17 to 18/8/7, from 36/25/20 to 15/8/11, from 13/8/8 to 10/5/5, respectively. Thus, the size, depth  and the complexity of the circuits are largely reduced. We next experimentally demonstrate the simplified schemes on IBM quantum computer. The evaluations indicate that the simplified schemes have a good fidelity above 0.9 for all the cases.

The present paper will be organized as follows, Section \ref{sec2} presents the simplified quantum circuits for teleporting an arbitrary single-qubit message via various entangled states in detail.  Experimentally demonstrations and the evaluation of the simplified schemes on IBM quantum computer are given in Section \ref{sec3}.  Finally, a conclusion is given in Section \ref{sec4}.

\section{Simplified quantum circuits for quantum teleportation} \label{sec2}

\begin{figure}[htbp]
	\centering
	\includegraphics[width=0.9\linewidth]{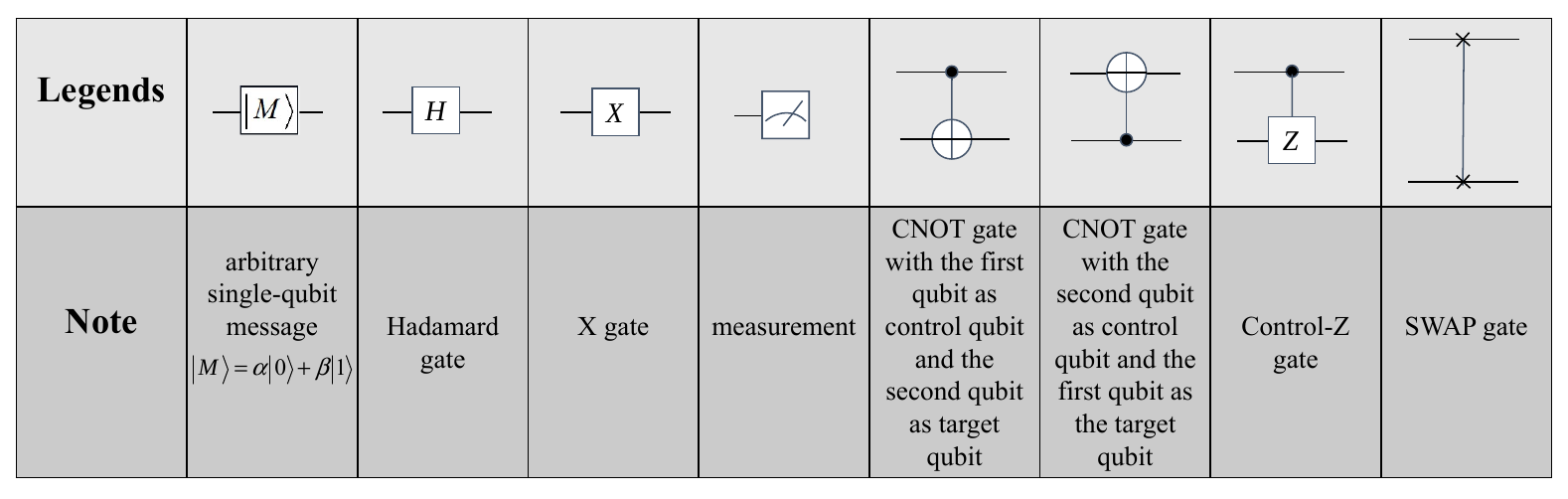}
	\caption{Legend description.\upcite{book}}
	\label{fig:legends}
\end{figure}

The quantum circuit model is the dominant paradigm for implementing complex quantum information processing task. Lowing the cost, depth, and size in the quantum circuit is better from the perspective of gate errors, noise resiliency, and lose state. Therefore, it is important to simplify quantum circuit as small as possible and as shallow as possible.
\textbf{Figure \ref{fig:legends}} shows some legends which will be employed to construct the simplified quantum circuits.

\subsection{Simplified quantum circuit for three-qubit-GHZ-based quantum teleportation}

We suppose that the single-qubit quantum message we need to teleport can be written as
\begin{eqnarray}\label{1}
	|M\rangle  = \alpha |0\rangle  + \beta |1\rangle,
\end{eqnarray}
where ${|\alpha|^2} + |\beta|^2 = 1$. To teleport the message, Alice, Bob, and Charlie need to share a GHZ state channel
\begin{eqnarray}\label{2}
	|\Phi\rangle  = \frac{1}{{\sqrt 2 }}(|000\rangle  + |111\rangle).
\end{eqnarray}

\begin{figure}[htbp]
	\centering
	\includegraphics[width=0.5\linewidth]{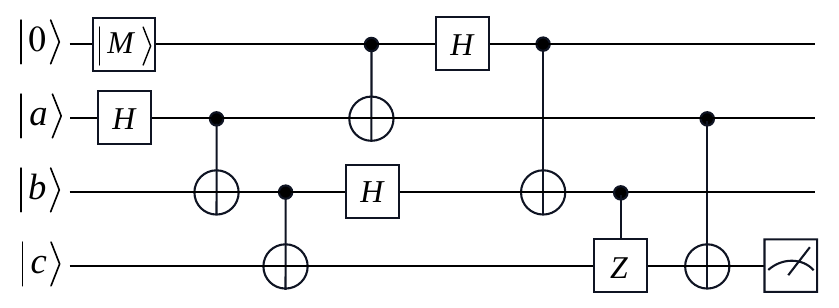}
	\caption{The original GHZ-based quantum teleportation.\upcite{old} $|M \rangle =\alpha|0\rangle + \beta|1\rangle$, and it can be implemented by employing one fundamental single-qubit gate. Here $|a \rangle$, $|b\rangle$, and $|c\rangle$ refer to $|0\rangle$.}
	\label{GHZ-old}
\end{figure}

Here the first, the second, and the third qubits belong to Alice, Bob, and  Charlie, respectively. The quantum circuit shown in  \textbf{Figure \ref{GHZ-old}} for teleporting the message $|M\rangle  = \alpha|0\rangle  + \beta|1\rangle$ from Alice to Bob can be further simplified and compressed.

\begin{figure}[htbp]
	\centering
	\includegraphics[width=0.5\linewidth]{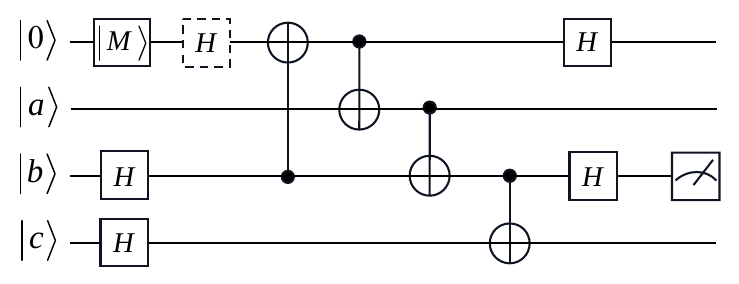}
	\caption{The simplified GHZ-based quantum teleportation. The single-qubit message can be implemented by employing three single-qubit rotations. Therefore, the Hadamard gate $H$ in dashed box can be absorbed by the neighborhood $|M\rangle$. Here $|a \rangle$, $|b\rangle$, and $|c\rangle$ refer to $|0\rangle$.}
	\label{GHZ-new}
\end{figure}

As shown in \textbf{Figure \ref{GHZ-new}}, by using the following tricks
\begin{eqnarray}\label{3}
	\text{CNOT}_3^2  \cdot  \text{CNOT}_2^1  \cdot  \text{CNOT}_3^2 = \text{CNOT}_2^1  \cdot  \text{CNOT}_3^1,
\end{eqnarray}
\begin{eqnarray}\label{4}
	\text{CNOT}_3^2 \cdot \text{CNOT}_1^3  \cdot  \text{CNOT}_3^2 = \text{CNOT}_1^3  \cdot  \text{CNOT}_1^2,
\end{eqnarray}
\begin{eqnarray}\label{5}
	(I \otimes H)  \cdot  \text{CNOT}_2^1  \cdot  (I \otimes H) = (H \otimes I)  \cdot  \text{CNOT}_1^2  \cdot  (H \otimes I),
\end{eqnarray}

The gate-count/cost/depth of \textbf{Figure \ref{GHZ-old}} is reduced from 10/6/8  to 9/4/6. Here CNOT$_i^j$ is a CNOT gate with the $i$-th and the $j$-th qubits encoded as the control and the target qubits, respectively. CNOT gate is one of the most popular gates and has been experimentally demonstrated in various systems.\upcite{CNOT1,CNOT2,CNOT3,CNOT4,CNOT5,CNOT6} The matrix forms of CNOT$_1^2$ and CNOT$_2^1$ are given by\upcite{book}
\begin{eqnarray}
	\text{CNOT}_1^2 = \left(
	\begin{array}{cccc}
		1 & 0 & 0 & 0\\
		0 & 1 & 0 & 0\\
		0 & 0 & 0 & 1\\
		0 & 0 & 1 & 0\\
	\end{array}\right),\;\;\;\;
	\text{CNOT}_2^1 = \left(
	\begin{array}{cccc}
		1 & 0 & 0 & 0\\
		0 & 0 & 0 & 1\\
		0 & 0 & 1 & 0\\
		0 & 1 & 0 & 0\\
	\end{array}\right),
\end{eqnarray}
in the $\{|00\rangle,\; |01\rangle,\; |10\rangle,\; |11\rangle\}$ basis. The operation of the Hadamard gate $H$ is given by\upcite{book}
\begin{eqnarray}\label{6}
	H = \frac{1}{\sqrt{2}}\left(
	\begin{array}{cc}
		1 & 1 \\
		1 & -1 \\
	\end{array}\right),
\end{eqnarray}
in the $\{|0\rangle, |1\rangle\}$ basis.

Alice, Bob, and Charlie applied a Hadamard operation on the first, third, and fourth qubits, respectively. Then state of the composite quantum system is evolved from $|\psi_0\rangle$ to $|\psi_1\rangle$. Here
\begin{eqnarray}\label{7}
	|\psi_0\rangle = (\alpha|0\rangle  + \beta |1\rangle)_1|000\rangle_{234},
\end{eqnarray}
\begin{equation}\label{8}
	\begin{aligned}
		|\psi_1\rangle=\frac{1}{2\sqrt 2}&[\alpha(|0\rangle +|1\rangle)_1 (|000\rangle + |001\rangle + |010\rangle + |011\rangle)_{234}\\
		+&\beta(|0\rangle - |1\rangle)_1(|000\rangle  + |001\rangle + |010\rangle +|011\rangle)_{234}].
	\end{aligned}
\end{equation}

Subsequently, operations $\text{CNOT}_3^1$, $\text{CNOT}_1^2$, $\text{CNOT}_2^3$, and $\text{CNOT}_3^4$ are performed on the system in succession. These CNOT gates transform $|\psi_1\rangle$ into
\begin{eqnarray}\label{9}
	\begin{aligned}
		|\psi_2\rangle = \frac{1}{2\sqrt 2}
		[  & \alpha|00\rangle_{24}(|00\rangle + |01\rangle)_{13} + \alpha|01\rangle_{24}(|00\rangle + |01\rangle)_{13}\\
		+&\alpha|10\rangle_{24}(|10\rangle + |11\rangle)_{13}+ \alpha|11\rangle_{24}(|10\rangle + |11\rangle)_{13}\\ +&\beta |00\rangle_{24}(|00\rangle - |01\rangle)_{13} + \beta |01\rangle_{24}(|00\rangle- |01\rangle)_{13}\\
		+& \beta |10\rangle_{24}(|10\rangle - |11\rangle)_{13} + \beta |11\rangle_{24}(|10\rangle - |11\rangle)_{13}].
	\end{aligned}
\end{eqnarray}

Finally, the two Hadamard gates make $|\psi_2\rangle$ become
\begin{eqnarray}\label{10}
	\begin{aligned}
		|\psi_3\rangle =\frac{1}{{4\sqrt 2}}[ &|000\rangle _{124}(\alpha|0\rangle + \beta|1\rangle)_3 + |001\rangle_{124}(\alpha|0\rangle + \beta|1\rangle)_3\\
		+&|010\rangle_{124}(\alpha|0\rangle + \beta|1\rangle)_3 + |011\rangle_{124}(\alpha|0\rangle + \beta|1\rangle)_3\\
		+&|100\rangle_{124}(\alpha|0\rangle + \beta|1\rangle)_3 - |101\rangle_{124}(\alpha|0\rangle + \beta|1\rangle)_3\\
		+&|110\rangle_{124}(\alpha|0\rangle + \beta|1\rangle)_3 - |111\rangle_{124}(\alpha|0\rangle + \beta|1\rangle)_3].
	\end{aligned}
\end{eqnarray}

Based on Equations (\ref{7}-\ref{10}), we can find that the simplified scheme shown in \textbf{Figure \ref{GHZ-new}} teleports a single-qubit message $|M\rangle  = \alpha|0\rangle  + \beta|1\rangle$  successfully without any feed-forward recover operation.

\subsection{Simplified quantum circuit for two-qubit-cluster-based quantum teleportation}

\begin{figure}[htbp]
	\centering
	\includegraphics[width=0.5\linewidth]{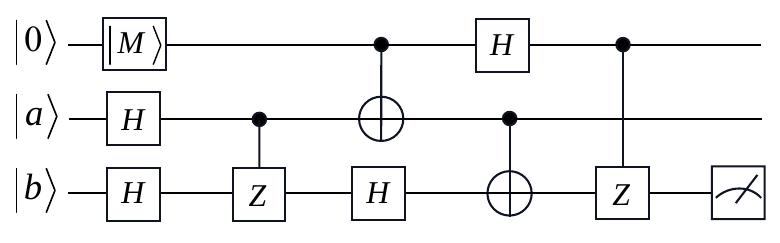}
	\caption{The original two-qubit-cluster-based quantum teleportation.\upcite{old} Here $|a\rangle$ and $|b\rangle$ refer to $|0\rangle$.}
	\label{two-qubit-cluster-old}
\end{figure}

\begin{figure}[htbp]
	\centering
	\includegraphics[width=0.5\linewidth]{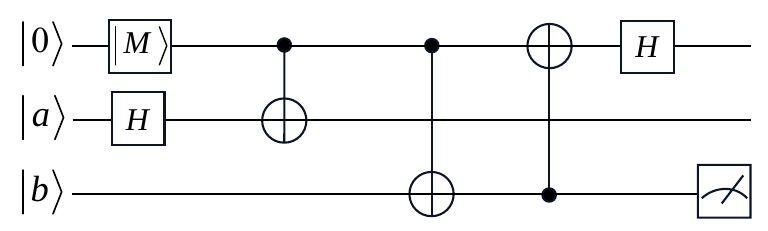}
	\caption{The simplified two-qubit-cluster-based quantum teleportation. Here $|a\rangle$ and $|b\rangle$ refer to $|0\rangle$.}
	\label{two-qubit-cluster-new}
\end{figure}

\textbf{Figure \ref{two-qubit-cluster-old}} shows a scheme for teleporting the normalization message $|M\rangle = \alpha|0\rangle_1 + \beta |1\rangle_1$ from Alice to Bob in two-qubit cluster state channel $\frac{1}{2}(|00\rangle +|01\rangle + |10\rangle - |11\rangle)_{23}$. Here Alice holds the first and second qubits, Bob holds the third qubit.

By employing the following tricks
\begin{eqnarray}\label{11}
	\text{CZ} = (I \otimes H) \cdot \text{CNOT}  \cdot  (I \otimes H),
\end{eqnarray}
\begin{eqnarray}\label{12}
	{\text{CNOT}_2^3 \cdot \text{CNOT}_1^2 \cdot \text{CNOT}_2^3 = \text{CNOT}_1^2 \cdot \text{CNOT}_1^3},
\end{eqnarray}
\begin{eqnarray}\label{13}
	(H \otimes H) \cdot \text{CNOT}_1^2 \cdot (H \otimes H) = \text{CNOT}_2^1.
\end{eqnarray}
Here the matrix form of the basic CZ gate is given by\upcite{book}
\begin{eqnarray}
	\text{CZ} =\left(
	\begin{array}{cccc}
		1 & 0 & 0 & 0\\
		0 & 1 & 0 & 0\\
		0 & 0 & 1 & 0\\
		0 & 0 & 0 & -1\\
	\end{array}\right),
\end{eqnarray}
in the $\{|00\rangle, |01\rangle, |10\rangle, |11\rangle\}$ basis.

\textbf{Figure \ref{two-qubit-cluster-old}} can be simplified as \textbf{Figure \ref{two-qubit-cluster-new}}, that is, the gate-count/cost/depth of the quantum circuit is decreased from 9/4/5 to 6/3/5.

As depicted in \textbf{Figure \ref{two-qubit-cluster-new}}, first, Alice performs a Hadamard gate on the second qubit, and it makes the system evolve from $|\varphi_0\rangle$ to $|\varphi_1\rangle$. Here
\begin{eqnarray}\label{14}
	|\varphi_0\rangle = (\alpha|0\rangle + \beta |1\rangle)_1|00\rangle_{23},
\end{eqnarray}
\begin{eqnarray}\label{15}
	|\varphi_1\rangle=\frac{1}{\sqrt2}(\alpha|000\rangle + \alpha|010\rangle + \beta|100\rangle + \beta|110\rangle)_{123}.
\end{eqnarray}

Next, $\text{CNOT}_1^2$, $\text{CNOT}_1^3$, and  $\text{CNOT}_3^1$ are performed in succession, changing the state $|\varphi_1\rangle$ into
\begin{eqnarray}
	|\varphi _2\rangle = \frac{1}{\sqrt 2}(\alpha|{000}\rangle + \alpha|{010}\rangle + \beta|{011}\rangle +\beta|{001}\rangle)_{123}.
\end{eqnarray}

Finally, a Hadamard operation is applied, resulting in
\begin{eqnarray}\label{16}
	\begin{aligned}
		|\varphi_3\rangle=\frac{1}{2\sqrt2}[&|00\rangle_{12}(\alpha|0\rangle + \beta|1\rangle)_3 + |10\rangle_{12}(\alpha|0\rangle + \beta|1\rangle)_3\\
		+&|01\rangle_{12}(\alpha|0\rangle + \beta|1\rangle)_3 + |11\rangle_{12}(\alpha|0\rangle + \beta|1\rangle)_3].
	\end{aligned}
\end{eqnarray}

Putting all the pieces together, we can find that the simplified quantum circuit shown in \textbf{Figure \ref{two-qubit-cluster-new}} can efficiently teleport the single-qubit quantum message $|M\rangle = \alpha|0\rangle + \beta|1\rangle$ without any feed-forward recover operations.

\subsection{Simplified quantum circuit for three-qubit-cluster-based quantum teleportation}

\begin{figure}[htbp]
	\centering
	\includegraphics[width=0.55\linewidth]{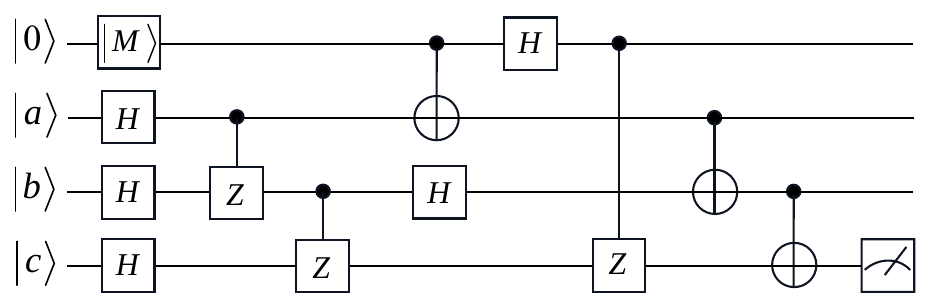}
	\caption{The original three-qubit-cluster-based quantum teleportation.\upcite{old} Here $|a\rangle$, $|b\rangle$, and $|c\rangle$ refer to $|0\rangle$.}
	\label{three-qubit-cluster-old}
\end{figure}

\begin{figure}[htbp]
	\centering
	\includegraphics[width=0.45\linewidth]{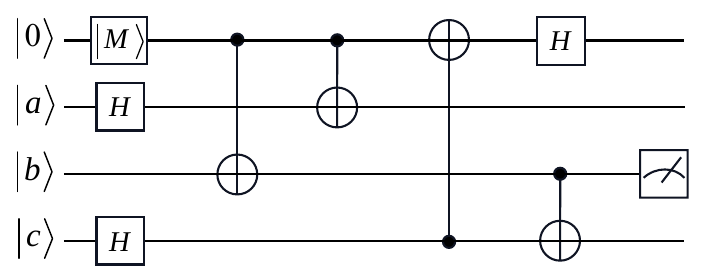}
	\caption{The simplified three-qubit-cluster-based quantum teleportation. Here $|a\rangle$, $|b\rangle$, and $|c\rangle$ refer to $|0\rangle$.}
	\label{three-qubit-cluster-new}
\end{figure}

The gate-count, cost, and depth of the quantum circuit for implementing quantum teleportation in three-qubit cluster channel ($\frac{1}{2\sqrt2}(|000\rangle + |001\rangle + |010\rangle - |011\rangle + |100\rangle + |101\rangle - |110\rangle + |111\rangle$), see \textbf{Figure \ref{three-qubit-cluster-old}}, can also be further decreased.
To simplify this circuit, we employ the relations between SWAP gate acting on the $i$-th and $j$-th qubits and CNOT gate, i.e.,
\begin{eqnarray}\label{17}
	\begin{aligned}
		\text{CNOT}_i^j \cdot \text{CNOT}_j^i \cdot \text{CNOT}_i^j = \text{SWAP}=\text{CNOT}_j^i \cdot \text{CNOT}_i^j \cdot \text{CNOT}_j^i.
	\end{aligned}
\end{eqnarray}

As shown in \textbf{Figure \ref{three-qubit-cluster-new}}, the initial state of the composite system is given by
\begin{eqnarray}\label{18}
	|\phi_0\rangle = (\alpha|0\rangle + \beta |1\rangle)_1|000\rangle_{234}.
\end{eqnarray}

First, the leftmost two Hadamard operations make $|\phi_0\rangle$ be changed into
\begin{eqnarray}\label{19}
	\begin{aligned}
		|\phi_1\rangle = \frac{1}{2}(&\alpha|0000\rangle + \alpha|0001\rangle + \alpha|0100\rangle + \alpha|0101\rangle\\
		+& \beta|1000\rangle + \beta |1001\rangle + \beta |1100\rangle + \beta |1101\rangle)_{1234}.
	\end{aligned}
\end{eqnarray}

Next, $\text{CNOT}_1^3$, $\text{CNOT}_1^2$, $\text{CNOT}_4^1$, and $\text{CNOT}_3^4$ are applied in succession, transforming $|\phi_1\rangle $ into
%
%

%
%

%
%

%
\begin{eqnarray}
	\begin{aligned}
		|\phi _2\rangle = \frac{1}{2}(&\alpha|{0000}\rangle + \alpha|{1001} \rangle + \alpha|{0100}\rangle + \alpha| {1101}\rangle\\
		+& \beta|{1111} \rangle + \beta|{0110}\rangle + \beta|{1011}\rangle + \beta|{0010} \rangle)_{1234}.
	\end{aligned}
\end{eqnarray}

At last, a Hadamard gate is performed on the first qubit, and then $|\phi _2\rangle$ becomes
\begin{eqnarray}\label{20}
	\begin{aligned}
		|\phi_3\rangle =\frac{1}{2\sqrt2}[& |000\rangle_{124}(\alpha|0\rangle + \beta|1\rangle)_3 + |100\rangle_{124}(\alpha|0\rangle + \beta|1\rangle)_3\\
		+&|001\rangle_{124}(\alpha|0\rangle + \beta|1\rangle)_3
		-|101\rangle_{124}(\alpha|0\rangle + \beta|1\rangle)_3\\
		+&|010\rangle_{124}(\alpha|0\rangle + \beta|1\rangle)_3
		+|110\rangle_{124}(\alpha|0\rangle+\beta|1\rangle)_3\\
		+&|011\rangle_{124}(\alpha|0\rangle + \beta|1\rangle)_3
		-|111\rangle_{124}(\alpha|0\rangle + \beta|1\rangle)_3].
	\end{aligned}
\end{eqnarray}

Based on Equations (\ref{18}-\ref{20}), we can find that the quantum circuit shown in \textbf{Figure \ref{three-qubit-cluster-new}} can efficiently complete a quantum teleportation without any feed-forward recover operations. That is, the gate-count/cost/depth is reduced from 12/6/7 to 8/4/5.

\subsection{Simplified quantum circuit for Brown-based quantum teleportation}

Brown state was first proposed by Brown et al.\upcite{21} in 2005 to teleport an arbitrary single-qubit message.  The Brown state can be written as
\begin{eqnarray}\label{21}
	\begin{aligned}
		|\Psi_{\text{Brown}}\rangle = \frac{1}{2}(|001\rangle|\Phi^-\rangle + |010\rangle|\Psi^-\rangle + |100\rangle|\Phi^+\rangle + |001\rangle|\Psi^+\rangle),
	\end{aligned}
\end{eqnarray}
where
\begin{eqnarray}\label{22}
	\begin{aligned}
		|\Psi^\pm\rangle  = \frac{1}{\sqrt2}(|00\rangle \pm |11\rangle),\quad
		|\Phi^\pm\rangle  = \frac{1}{\sqrt2}(|10\rangle \pm |10\rangle).
	\end{aligned}
\end{eqnarray}

\begin{figure}[htbp]
	\centering
	\includegraphics[width=0.75\linewidth]{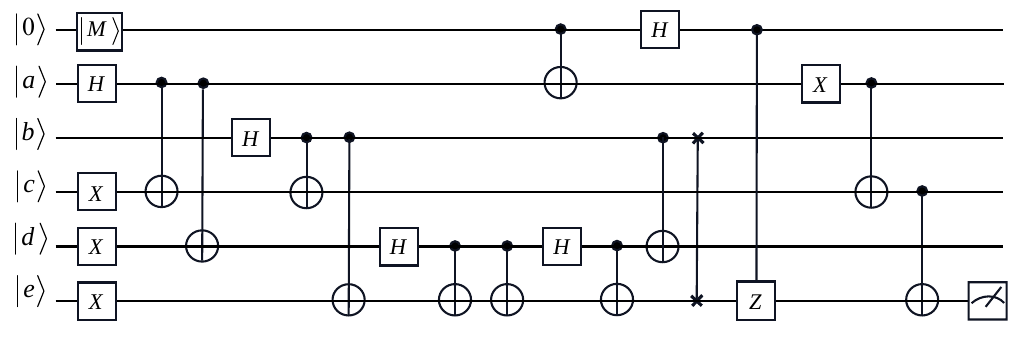}
	\caption{The original Brown-based quantum teleportation.\upcite{old} Here $|a\rangle$, $|b\rangle$, $|c\rangle$, $|d\rangle$, and $|e\rangle$ refer to $|0\rangle$.}
	\label{Brown-old}
\end{figure}

\begin{figure}[htbp]
	\centering
	\includegraphics[width=0.75\linewidth]{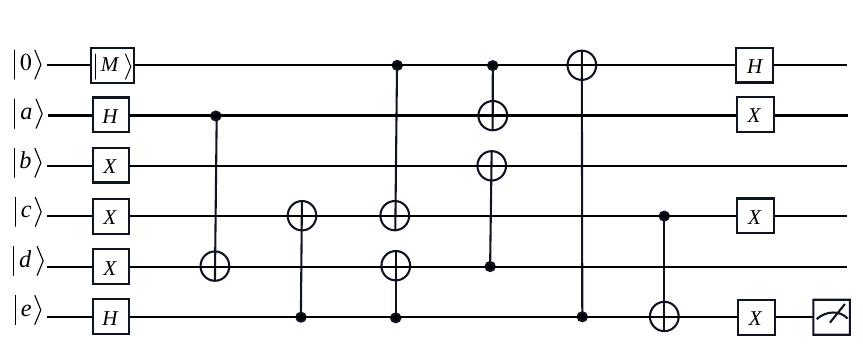}
	\caption{The simplified Brown-based quantum teleportation. Here $|a\rangle$, $|b\rangle$, $|c\rangle$, $|d\rangle$, and $|e\rangle$ refer to $|0\rangle$.}
	\label{Brown-new}
\end{figure}

\textbf{Figure \ref{Brown-old}} shows a quantum circuit for Brown-based quantum teleportation, and its gate-count/cost/depth can be decreased from 25/15/17 to 18/8/7 by taking the following equivalence relations
\begin{eqnarray}\label{23}
	(X \otimes I) \cdot \text{CNOT} = \text{CNOT} \cdot(X \otimes X),
\end{eqnarray}
\begin{eqnarray}\label{24}
	\text{CNOT}_2^4 \cdot \text{CNOT}_1^2 \cdot \text{CNOT}_2^4 = \text{CNOT}_1^2 \cdot \text{CNOT}_1^4,
\end{eqnarray}
\begin{eqnarray}\label{25}
	\text{CNOT}_3^4 \cdot \text{CNOT}_1^3 \cdot \text{CNOT}_3^4 = \text{CNOT}_1^3 \cdot \text{CNOT}_1^4.
\end{eqnarray}
Here the transformation of  bit-flip gate $X$ can be expressed as\upcite{book}
\begin{eqnarray}
	X = \left(
	\begin{array}{cc}
		0 & 1 \\
		1 & 0 \\
	\end{array}\right),
\end{eqnarray}
in the $\{|0\rangle, |1\rangle\}$ basis.

As shown in \textbf{Figure \ref{Brown-new}}, we first apply the leftmost two Hadamard gates and three $X$ gates on the composite system. These operations will result in the state
\begin{eqnarray}\label{26}
	\begin{aligned}
		|\chi_1\rangle  = \frac{1}{2}[&\alpha(|001110\rangle  + |001111\rangle  + |011110\rangle  + |011111\rangle)_{123456}\\
		+ &\beta (|101110\rangle  + |101111\rangle  + |111110\rangle  + |111111\rangle)_{123456}].
	\end{aligned}
\end{eqnarray}

Then $\text{CNOT}_2^5$ and $\text{CNOT}_6^4$ are applied to $|\chi_1\rangle$, and they transform $|\chi_1\rangle$ into
\begin{eqnarray}
	\begin{aligned}
		|\chi _2\rangle = \frac{(\alpha|0\rangle + \beta|1\rangle )_1}{2}(| {01110}\rangle + |{11100}\rangle + |{01011}\rangle + | {11001}\rangle)_{23456}.
	\end{aligned}
\end{eqnarray}

After $\text{CNOT}_1^4$, $\text{CNOT}_6^5$, $\text{CNOT}_1^2$, $\text{CNOT}_5^3$, $\text{CNOT}_6^1$, and $\text{CNOT}_4^6$ are performed, the system becomes
\begin{eqnarray}\label{27}
	\begin{aligned}
		|\chi_3\rangle=\frac{1}{2\sqrt2}[&|00011\rangle_{12345}(\alpha|1\rangle + \beta|0\rangle)_6
		+ |01110\rangle_{12345}(\alpha|1\rangle + \beta|0\rangle)_6\\
		+&|10011\rangle_{12345}(\alpha|1\rangle + \beta|0\rangle)_6
		+ |11110\rangle_{12345}(\alpha|1\rangle + \beta|0\rangle)_6\\
		+&|00100\rangle_{12345}(\alpha|1\rangle + \beta|0\rangle)_6
		+ |01001\rangle_{12345}(\alpha|1\rangle + \beta|0\rangle)_6\\
		-&|10100\rangle_{12345}(\alpha|1\rangle + \beta|0\rangle)_6
		- |11001\rangle_{12345}(\alpha|1\rangle + \beta|0\rangle)_6].
	\end{aligned}
\end{eqnarray}

Finally, the rightmost Hadamard gate and X gates make $|\chi_3\rangle$ be changed into
\begin{eqnarray}\label{28}
	\begin{aligned}
		|\chi_4\rangle= \frac{1}{2\sqrt2}[& |01001\rangle_{12345}(\alpha|0\rangle + \beta|1\rangle)_6
		+  |00100\rangle_{12345}(\alpha|0\rangle + \beta|1\rangle)_6\\
		+ &|11001\rangle_{12345}(\alpha|0\rangle + \beta|1\rangle)_6
		+  |10100\rangle_{12345}(\alpha|0\rangle + \beta|1\rangle)_6\\
		+ &|01110\rangle_{12345}(\alpha|0\rangle + \beta|1\rangle)_6
		+  |00011\rangle_{12345}(\alpha|0\rangle + \beta|1\rangle)_6\\
		- &|11110\rangle_{12345}(\alpha|0\rangle + \beta|1\rangle)_6
		-  |10011\rangle_{12345}(\alpha|0\rangle + \beta|1\rangle)_6].
	\end{aligned}
\end{eqnarray}

Combining Equations (\ref{26}-\ref{28}), one can see that the simplified scheme shown in \textbf{Figure \ref{Brown-new}} successfully teleport the single-qubit message $|M\rangle = \alpha|0\rangle + \beta|1\rangle$ from the first qubit to the sixth qubit without any recover operations.

\subsection{Simplified quantum circuit for Borras-based quantum teleportation}

In 2007, Borras et al.\upcite{22} introduced a genuinely entangled six-qubit state, which is given by
\begin{eqnarray}\label{29}
	\begin{aligned}
		|\Psi\rangle =\frac{1}{4}[& |000\rangle (|0\rangle  |\Psi^+\rangle  + |1\rangle |\Phi^+\rangle) + |001\rangle  (|0\rangle |\Phi^-\rangle - |1\rangle |\Psi^-\rangle)\\
		+&|010\rangle (|0\rangle  |\Phi^+\rangle  - |1\rangle |\Psi^+\rangle) + |011\rangle  (|0\rangle |\Psi^-\rangle + |1\rangle |\Phi^-\rangle)\\
		-&|100\rangle (|0\rangle  |\Phi^-\rangle  + |1\rangle |\Psi^-\rangle) - |101\rangle  (|0\rangle |\Psi^+\rangle - |1\rangle |\Phi^+\rangle)\\
		+&|110\rangle (|0\rangle  |\Psi^-\rangle  - |1\rangle |\Phi^-\rangle) + |111\rangle  (|0\rangle |\Psi^+\rangle + |1\rangle |\Psi^+\rangle)].
	\end{aligned}
\end{eqnarray}

\begin{figure}[htbp]
	\centering
	\includegraphics[width=0.96\linewidth]{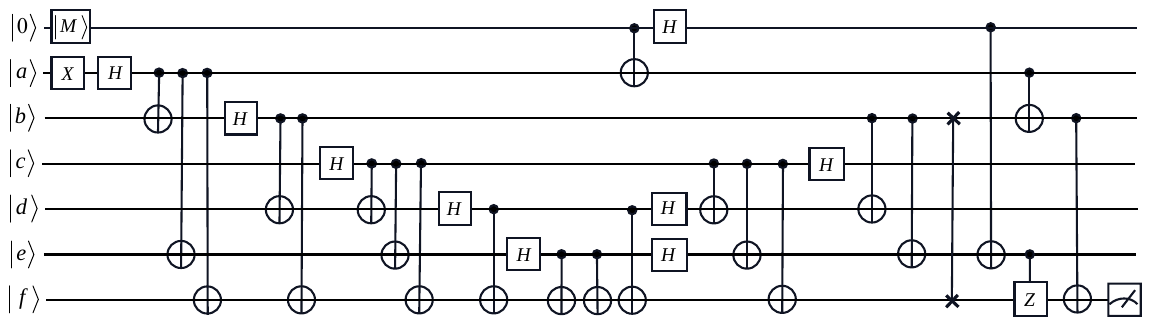}
	\caption{The original Borras-based quantum teleportation.\upcite{old} Here $|a\rangle$, $|b\rangle$, $|c\rangle$, $|d\rangle$, $|e\rangle$, and $|f\rangle$ refer to $|0\rangle$.}
	\label{Borras-old}
\end{figure}

As shown in \textbf{Figure \ref{Borras-old}}, the quantum circuit for teleporting a single-qubit message among Borras state has been designed by using 25 CNOT gates and 11 single-qubit fundamental gates.  Fortunately, the gate-count/cost/depth of this elaborate circuit can be reduced from 36/25/20 to 15/8/11 by introducing the following relations:
\begin{eqnarray}\label{30}
	\text{CNOT}_2^7 \cdot \text{CNOT}_1^2 \cdot \text{CNOT}_2^7 =\text{CNOT}_1^7 \cdot \text{CNOT}_1^2.
\end{eqnarray}

\begin{figure}[htbp]
	\centering
	\includegraphics[width=0.75\linewidth]{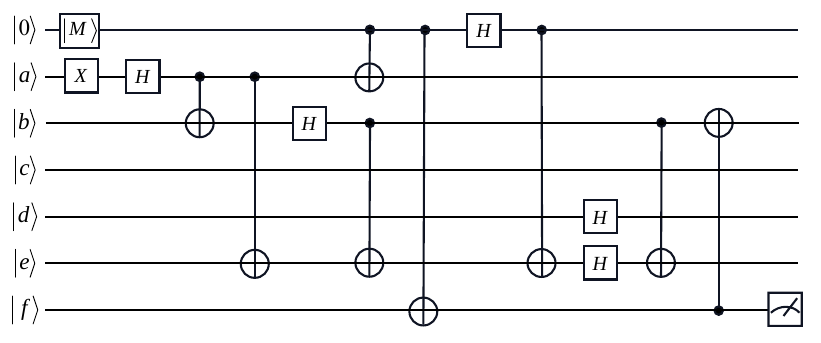}
	\caption{The simplified Borras-based quantum teleportation. Here $|a\rangle$, $|b\rangle$, $|c\rangle$, $|d\rangle$, $|e\rangle$, and $|f\rangle$ refer to $|0\rangle$.}
	\label{Borras-new}
\end{figure}

As shown in \textbf{Figure \ref{Borras-new}}, the leftmost operation $X$, $H$, CNOT$_2^3$, and CNOT$_2^6$ transform the state of the composite system from $|\omega_0\rangle$ to $|\omega_1\rangle$. Here
\begin{eqnarray}\label{31}
	|\omega_0\rangle = (\alpha|0\rangle_1 + \beta|1\rangle_1)|000000)_{234567},
\end{eqnarray}
\begin{eqnarray}\label{32}
	|\omega_1\rangle  = \frac{1}{\sqrt2}(\alpha|0000000\rangle - \alpha|0110010\rangle + \beta|1000000\rangle - \beta|1110010\rangle)_{1234567}.
\end{eqnarray}

Subsequently, the operations $H$ acting on the third qubit, CNOT$_1^2$, CNOT$_3^6$, CNOT$_1^7$, $H$ acting on the first qubit, and CNOT$_1^6$ are applied in succession, which will convert $|\omega_1\rangle$ into
\begin{eqnarray}\label{33}
	\begin{aligned}
		|\omega_2\rangle =\frac{1}{2\sqrt2}( &\alpha|0000000\rangle - \alpha|0100010\rangle + \alpha|0010010\rangle + \alpha|0110000\rangle \\
		+ &\alpha|1000010\rangle - \alpha|1100000\rangle + \alpha|1010000\rangle + \alpha|1110010\rangle \\
		+ & \beta|0100001\rangle - \beta |0000011\rangle + \beta |0110011\rangle + \beta |0010001\rangle\\
		- & \beta|1100011\rangle + \beta |1000001\rangle)- \beta |1110001\rangle - \beta |1010011\rangle)_{1234567}.
	\end{aligned}
\end{eqnarray}

Finally, after operations $H$ acting on the fifth and third qubits, CNOT$_3^6$, CNOT$_7^3$ are performed in succession, $|\omega_2\rangle$ becomes
\begin{eqnarray}\label{34}
	\begin{aligned}
		|\omega_3\rangle=\frac{1}{4\sqrt2}[&|000000\rangle(\alpha|0\rangle + \beta|1\rangle) - |010000\rangle(\alpha|0\rangle + \beta|1\rangle)\\
		+&|001001\rangle(\alpha|0\rangle + \beta|1\rangle) + |000010\rangle(\alpha|0\rangle + \beta|1\rangle)\\
		-&|010010\rangle(\alpha|0\rangle + \beta|1\rangle) + |011011\rangle(\alpha|0\rangle + \beta|1\rangle)\\
		+&|100000\rangle(\alpha|0\rangle + \beta|1\rangle) - |110000\rangle(\alpha|0\rangle + \beta|1\rangle)\\
		+&|101001\rangle(\alpha|0\rangle + \beta|1\rangle) + |111001\rangle(\alpha|0\rangle + \beta|1\rangle)\\
		+&|100010\rangle(\alpha|0\rangle + \beta|1\rangle) - |110010\rangle(\alpha|0\rangle + \beta|1\rangle)\\
		+&|101011\rangle(\alpha|0\rangle + \beta|1\rangle) + |111011\rangle(\alpha|0\rangle + \beta|1\rangle)\\
		+&|000001\rangle(\alpha|0\rangle + \beta|1\rangle) + |010001\rangle(\alpha|0\rangle + \beta|1\rangle)\\
		-&|001000\rangle(\alpha|0\rangle + \beta|1\rangle) + |011000\rangle(\alpha|0\rangle + \beta|1\rangle)\\
		+&|000011\rangle(\alpha|0\rangle + \beta|1\rangle) + |010011\rangle(\alpha|0\rangle + \beta|1\rangle)\\
		-&|001010\rangle(\alpha|0\rangle + \beta|1\rangle) + |011010\rangle(\alpha|0\rangle + \beta|1\rangle)\\
		-&|100001\rangle(\alpha|0\rangle + \beta|1\rangle) - |110001\rangle(\alpha|0\rangle + \beta|1\rangle)\\
		+&|101000\rangle(\alpha|0\rangle + \beta|1\rangle) - |111000\rangle(\alpha|0\rangle + \beta|1\rangle)\\
		-&|100011\rangle(\alpha|0\rangle + \beta|1\rangle) - |110011\rangle(\alpha|0\rangle + \beta|1\rangle)\\
		+&|101010\rangle(\alpha|0\rangle + \beta|1\rangle) - |111010\rangle(\alpha|0\rangle + \beta|1\rangle)\\
		+&|001011\rangle(\alpha|0\rangle+ \beta|1\rangle) + |011001\rangle(\alpha|0\rangle + \beta|1\rangle)]_{1234567}.
	\end{aligned}
\end{eqnarray}

Therefore, the simplified quantum circuit shown in \textbf{Figure \ref{Borras-new}} can successfully teleport a message $|M\rangle  = \alpha |0\rangle  + \beta |1\rangle$ from the first qubit to the seventh qubit without any feed-forward recover operations.

\subsection{Simplified quantum circuit for entanglement-swapping-based quantum teleportation}

\begin{figure}[htbp]
	\centering
	\includegraphics[width=0.6\linewidth]{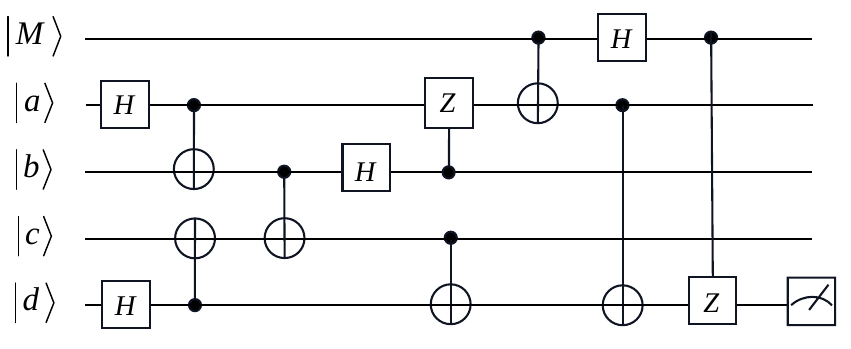}
	\caption{The original entanglement-swapping-based quantum teleportation.\upcite{old-} $|M\rangle  = \alpha |0\rangle  + \beta |1\rangle$. Here $|a\rangle$, $|b\rangle$, $|c\rangle$ and $|d\rangle$ refer to $|0\rangle$.}
	\label{entanglement-swapping-old}
\end{figure}

\begin{figure}[htbp]
	\centering
	\includegraphics[width=0.6\linewidth]{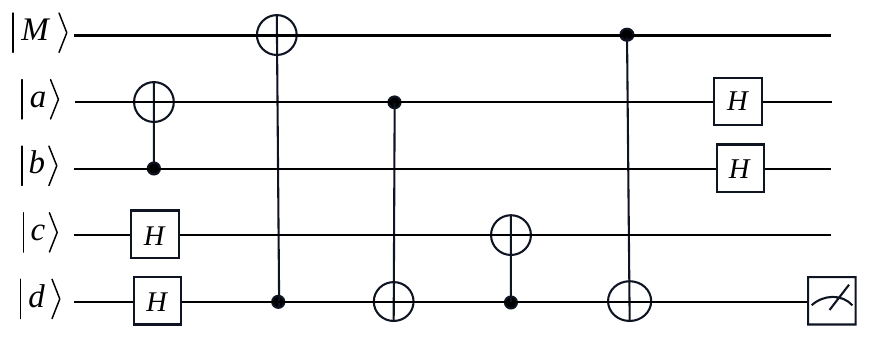}
	\caption{The simplified entanglement-swapping-based quantum teleportation. Here $|a\rangle$, $|b\rangle$, $|c\rangle$ and $|d\rangle$ refer to $|0\rangle$.}
	\label{entanglement-swapping-new}
\end{figure}

The teleportation of single-qubit message $|M\rangle  = \alpha |0\rangle  + \beta |1\rangle$ can also be achieved based on the entanglement swapping (see \textbf{Figure \ref{entanglement-swapping-old}}). As shown in \textbf{Figure \ref{entanglement-swapping-new}}, the gate-count/cost/depth can be further decreased from 13/8/8 to 10/5/5 by employing  the following relations
\begin{eqnarray}\label{35}
	\text{CNOT}_1^2 \cdot \text{CNOT}_2^3 \cdot \text{CNOT}_1^2 = \text{CNOT}_1^3 \cdot \text{CNOT}_2^3,
\end{eqnarray}
\begin{eqnarray}\label{36}
	\text{CNOT}_2^5 \cdot \text{CNOT}_5^1 \cdot \text{CNOT}_2^5 = \text{CNOT}_5^1 \cdot \text{CNOT}_2^1.
\end{eqnarray}

Firstly, the leftmost two Hadamard gates are applied, and they convert the state of the composite system from $|\eta_0\rangle$ into $|\eta_1\rangle$. Here
\begin{eqnarray}\label{37}
	|\eta_0\rangle  = (\alpha |0\rangle_1 + \beta|1\rangle_1)|0000\rangle_{2345},
\end{eqnarray}
\begin{eqnarray}\label{38}
	|\eta_1\rangle  = \frac{1}{2}(\alpha |0\rangle_1 + \beta |1\rangle_1)(|0000\rangle +|0001\rangle + |0010\rangle + |0011\rangle)_{2345}.
\end{eqnarray}

Subsequently, operations CNOT$_3^2$, CNOT$_5^1$, CNOT$_2^5$, CNOT$_5^4$, and CNOT$_1^5$ are performed in succession, and they will result in the state
\begin{eqnarray}\label{39}
	\begin{aligned}
		|\eta_2\rangle  = \frac{1}{2}&[\alpha|0\rangle(|0000\rangle + |0010\rangle) + \alpha |1\rangle (|0010\rangle + |0000\rangle)\\
		+& \beta |1\rangle(|0001\rangle +|0011\rangle)  + \beta  |0\rangle (|0011\rangle + |0001\rangle)]_{12345}.
	\end{aligned}
\end{eqnarray}

Lastly, the rightmost two Hadamard gates will transform $|\eta_2\rangle$ into
\begin{eqnarray}\label{40}
	\begin{aligned}
		|\eta_3\rangle = \frac{1}{4}(&|0000\rangle  + |0001\rangle  + |1001\rangle  + |1000\rangle
		+ |0100\rangle  + |0101\rangle\\
		+& |1101\rangle  + |1100\rangle
		+|0010\rangle  + |0011\rangle  + |1011\rangle  + |1010\rangle\\
		+& |0110\rangle  + |0111\rangle  + |1111\rangle  + |1110\rangle)_{1234}(\alpha |0\rangle + \beta|1\rangle)_5.
	\end{aligned}
\end{eqnarray}

Equations (\ref{37}-\ref{40}) indicate that the quantum circuit shown in \textbf{Figure \ref{entanglement-swapping-new}} successfully teleport a single-qubit message $|M\rangle  = \alpha |0\rangle  + \beta |1\rangle$ from the first qubit to the fifth qubit without any feed-forward recover operation.

\begin{figure}[htbp]
	\centering
	\includegraphics[width=0.7\linewidth]{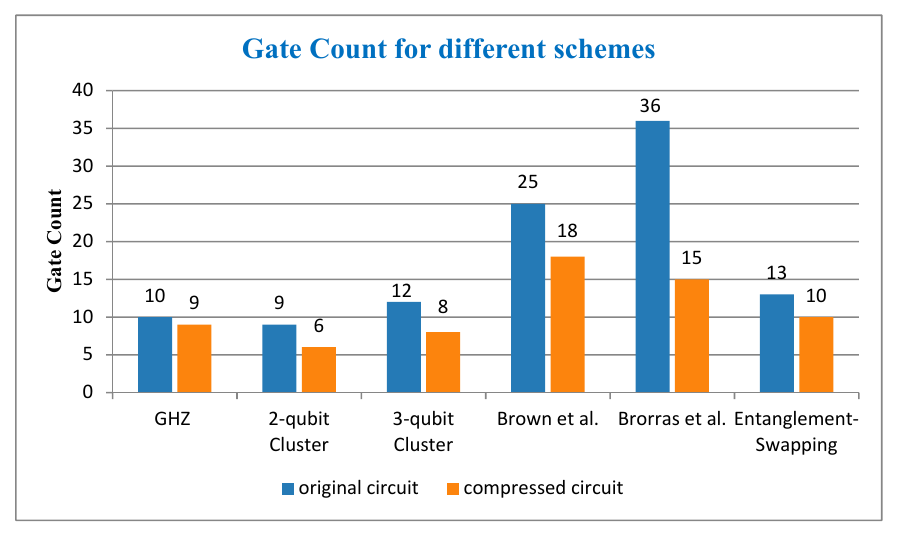}
	\caption{Gate Count for different schemes.}
	\label{fig:gate-count}
\end{figure}

\begin{figure}[htbp]
	\centering
	\includegraphics[width=0.7\linewidth]{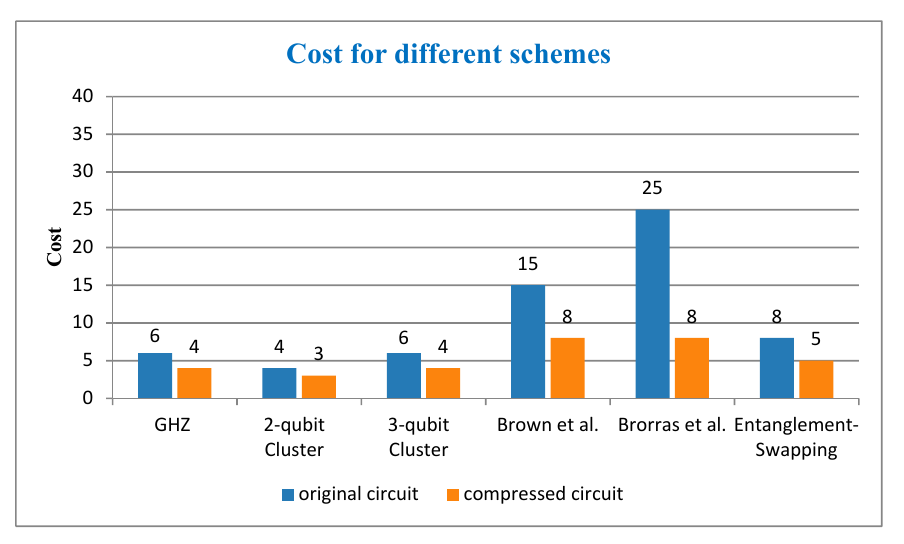}
	\caption{Cost for different schemes.}
	\label{fig:cost}
\end{figure}

\begin{figure}[htbp]
	\centering
	\includegraphics[width=0.7\linewidth]{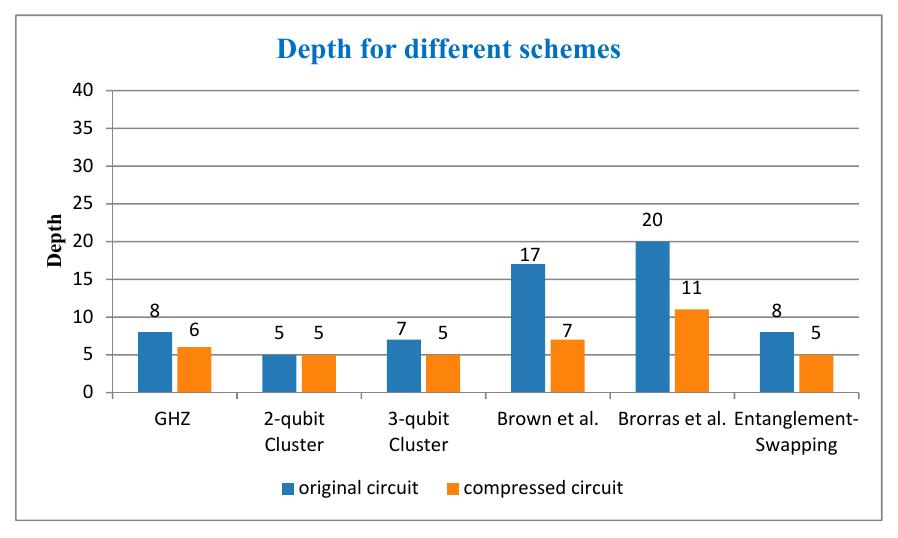}
	\caption{Depth for different schemes.}
	\label{fig:depth}
\end{figure}

Based on \textbf{Figures \ref{fig:gate-count}-\ref{fig:depth}}, we can find the quantum circuits for implementing quantum teleportation among different channels are largely simplified in terms of the gate-count, cost, and depth, which are better from the perspective of noise resiliency and resource-saving.  It is known that, quantum states
are fragile due to decoherence and imperfect operations, hence, we will evaluation of performance of the simplified and original schemes via IBM quantum computer in the next section.

\section{Experimental realizations with IBM quantum computers} \label{sec3}

As far as the current development is concerned, quantum cloud computing is one of the main ways of using quantum computing in recent years. IBM (International Business Machines Corporation) quantum experiment offers a cloud-based quantum computing platform. Using IBM quantum computer to realize and evaluate the performance of a quantum circuit is a common approach.\upcite{IBM1,quantum-teleportation1,experiment-cluster1,IBM4,IBM5,IBM6,IBM7,IBM8,IBM9,IBM10}

\begin{figure}[htbp]
	\centering
	\includegraphics[width=0.7\linewidth]{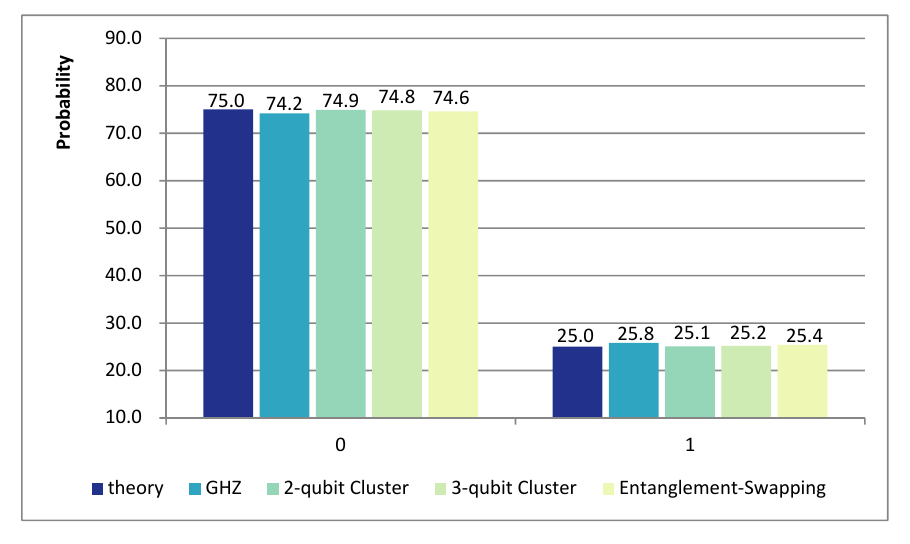}
	\caption{Demonstration of quantum teleportation of the state $|M\rangle  = \cos \frac{\pi }{6}|0\rangle + \sin \frac{\pi }{6}|1\rangle$.}
	\label{probability1}
\end{figure}

\begin{figure}[htbp]
	\centering
	\includegraphics[width=0.7\linewidth]{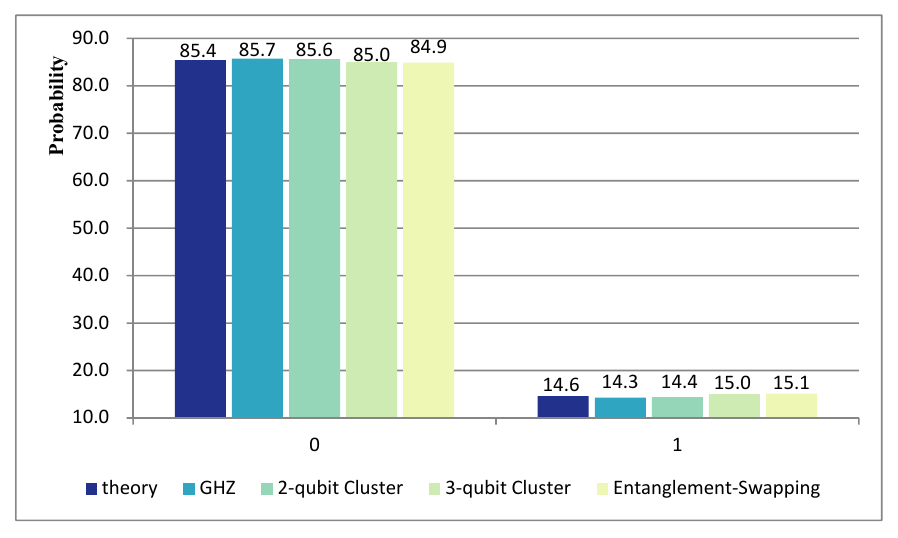}
	\caption{Demonstration of quantum teleportation of the state $|M\rangle = \cos\frac{\pi }{8}|0\rangle + \sin \frac{\pi }{8}|1\rangle $.}
	\label{probability2}
\end{figure}

Quantum state tomography\upcite{IBM1,quantum-state-tomogtaphy1,quantum-state-tomogtaphy2,quantum-state-tomogtaphy3} is a common method to characterize a quantum state for characterize the performance of the schemes.
We use the simulator called ``simulator extended stabilizer" (with 15360 shots for more accuracy and to reduce statistical errors) on IBM quantum experiment to tomography the single-qubit state $|M\rangle  = \cos \frac{\theta }{2}|0\rangle  + \sin \frac{\theta }{2}|1\rangle$ with $\theta=\frac{\pi }{3}$ (see \textbf{Figure \ref{probability1}}) and  $\theta=\frac{\pi }{4}$ (see \textbf{Figure \ref{probability2}}), respectively.

Fidelity is the measurement of the overlap between two density matrices of theoretical and experimental quantum states. Fidelity is given by\upcite{book}
\begin{eqnarray}\label{41}
	F(\rho ^T,\rho^E) = (\text{Tr}(\sqrt{\sqrt{\rho^T} \rho^E \sqrt{\rho^T}}))^2,
\end{eqnarray}
Here, $\rho^T$ and $\rho^E$ are the theoretical and experimental density matrix, respectively.

For the state $|M\rangle  = \cos \frac{\pi}{6}|0\rangle  + \sin \frac{\pi}{6}|1\rangle$, the density matrices of the presented simplified quantum circuits can be written as
\begin{eqnarray}\label{42}
	\rho^T =
	\left(
	\begin{array}{cc}
		0.750 & 0.4330\\
		0.4330 & 0.250\\
	\end{array}
	\right).
\end{eqnarray}

The density matrices of the simplified quantum circuits in experiment are
\begin{eqnarray}\label{43}
	\rho_{\text{GHZ}}^E = \left(\begin{array}{*{20}{c}}
		0.742 & 0.435\\
		0.435 & 0.258\\
	\end{array}
	\right),
\end{eqnarray}
\begin{eqnarray}\label{44}
	\rho_{\text{2-qubit-cluster}}^E = \left(\begin{array}{*{20}{c}}
		0.749          & 0.432 + 0.001i\\
		0.432 - 0.001i & 0.251         \\
	\end{array}
	\right),
\end{eqnarray}
\begin{eqnarray}\label{45}
	\rho_{\text{3-qubit-cluster}}^E = \left(\begin{array}{*{20}{c}}
		0.748            &  0.427 - 0.003i\\
		0.427 + 0.003i   &  0.252\\
	\end{array}
	\right),
\end{eqnarray}
\begin{eqnarray}\label{46}
	\rho_{\text{entanglement-swapping}}^E = \left(\begin{array}{*{20}{c}}
		0.474          &  0.436 - 0.002i\\
		0.436 + 0.002i & 0.253\\
	\end{array}
	\right).
\end{eqnarray}

\begin{figure}[htbp]
	\centering
	\includegraphics[width=0.7\linewidth]{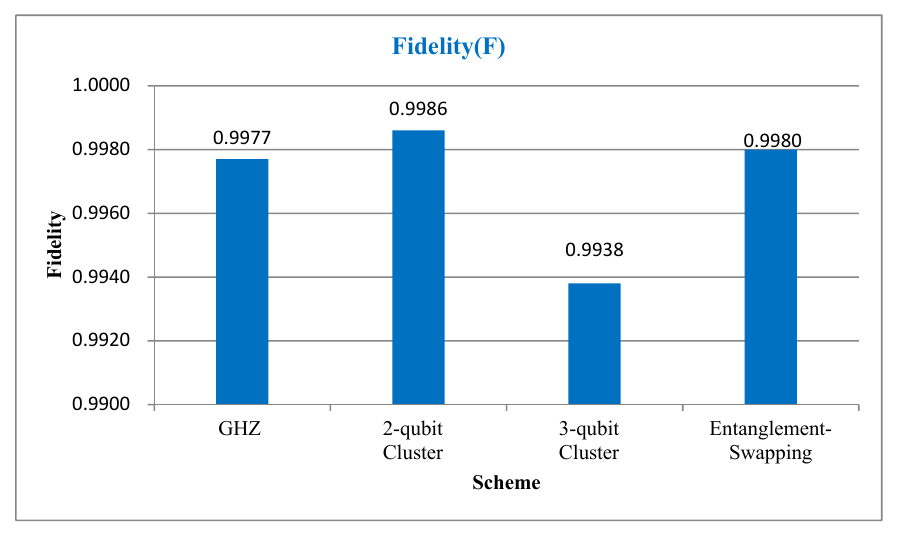}
	\caption{Fidelity between theoretical density matrix and experimental density matrices for the teleportation of $|M\rangle  = \cos \frac{\pi}{6}|0\rangle  + \sin \frac{\pi}{6}|1\rangle$ based on Equations (\ref{43}-\ref{46}).}
	\label{fig:fidelity-pi3}
\end{figure}

\textbf{Figure \ref{fig:fidelity-pi3}} shows the theoretical and experimental fidelity of the simplified GHZ-based, two-qubit-cluster-based, three-cluster-based, and entanglement-based quantum teleportation with $|M\rangle  = \cos \frac{\pi}{6}|0\rangle  + \sin \frac{\pi}{6}|1\rangle$.

For $|M\rangle=\cos\frac{\pi}{8}|0\rangle + \sin\frac{\pi}{8}|1\rangle$,  the theoretical density matrix $\rho_2^T$ can be expressed as
\begin{eqnarray}\label{47}
	\rho_2^T =  \left(\begin{array}{*{20}{c}}
		0.8536 & 0.3536\\
		0.3536 & 0.1464\\
	\end{array}\right).
\end{eqnarray}

The experimental density matrices can be expressed as
\begin{eqnarray}\label{48}
	\rho_{\text{GHZ}}^E =  \left(\begin{array}{*{20}{c}}
		0.857           & 0.345 - 0.002i\\
		0.345 + 0.002i  & 0.143 \\
	\end{array}\right),
\end{eqnarray}
\begin{eqnarray}\label{49}
	\rho_{\text{2-qubit-cluster}}^E =  \left(\begin{array}{*{20}{c}}
		0.856           &  0.349 + 0.008i\\
		0.349 - 0.008i  &  0.284\\
	\end{array}\right),
\end{eqnarray}
\begin{eqnarray}\label{50}
	\rho_{\text{3-qubit-cluster}}^E =  \left(\begin{array}{*{20}{c}}
		0.85            & 0.35 - 0.001i\\
		0.35 + 0.001i   & 0.15\\
	\end{array}\right),
\end{eqnarray}
\begin{eqnarray}\label{51}
	\rho_{\text{entanglement-swapping}}^E =  \left(\begin{array}{*{20}{c}}
		0.849           & 0.347 - 0.002i\\
		0.347 + 0.002i  & 0.151\\
	\end{array}\right).
\end{eqnarray}

\begin{figure}[htbp]
	\centering
	\includegraphics[width=0.7\linewidth]{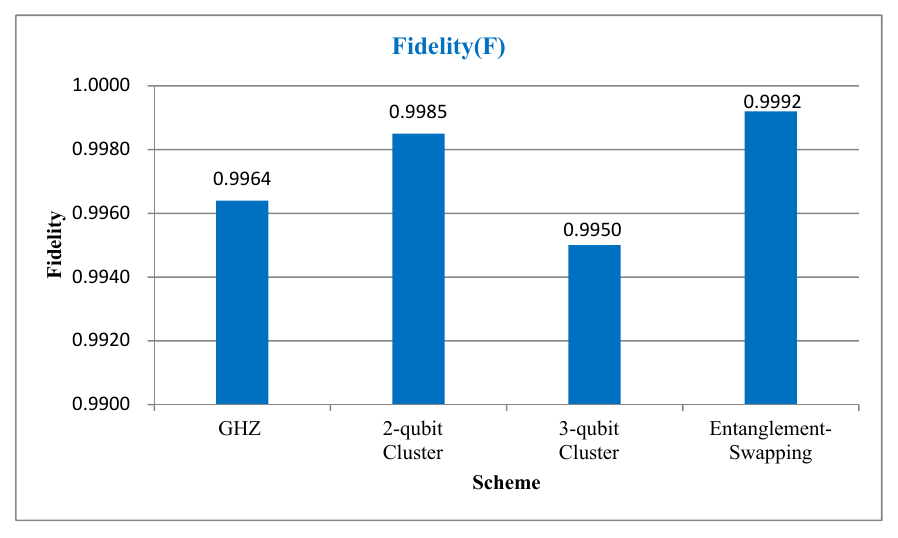}
	\caption{Fidelity between theoretical density matrix and experimental density matrices for teleportation of $|M\rangle=\cos\frac{\pi}{8}|0\rangle + \sin\frac{\pi}{8}|1\rangle$ based on Equations (\ref{47}-\ref{51}).}
	\label{fig:fidelity-pi4}
\end{figure}

\textbf{Figure \ref{fig:fidelity-pi4}} shows the theoretical and experimental fidelity of the simplified GHZ-based, two-qubit-cluster-based, three-qubit-cluster-based, and entanglement-based quantum teleportation with $|M\rangle  = \cos \frac{\pi}{8}|0\rangle  + \sin \frac{\pi}{8}|1\rangle$.

\section{Conclusion}\label{sec4}

It is crucial for designing a specific quantum information processing task as small as possible, as low-cost as possible, and as shallow as possible. In this paper, we simplified quantum circuits for implementing quantum teleportation via various entangled states, including GHZ state, two-qubit cluster state, three-qubit cluster state, Brown state,  Borras state, and entanglement swapping. By introducing some tricks, the gate count, cost and the circuit depth are dramatically deduced (see \textbf{Figure \ref{fig:gate-count}}, \textbf{Figure \ref{fig:cost}}, and \textbf{Figure \ref{fig:depth}}). The gate-count and cost of some schemes are even cut by more than half.  Our simplified quantum circuits will benefit both the quantum resource and the noise resistant.

IBM releases a five-qubit cloud experimental computing platform based on superconducting quantum bits. We have demonstrated the GHZ-based, two-qubit-cluster-based, three-cluster-based and the entanglement-swapping-based simplified schemes on IBM quantum computer successfully in \textbf{Figure \ref{probability1}} and \textbf{Figure \ref{probability2}}. \textbf{Figure \ref{fig:fidelity-pi3}} and \textbf{Figure \ref{fig:fidelity-pi4}} indicate that the fidelities of the simplified schemes are reasonably high.

\medskip
\section*{Acknowledgment} 
This work is supported by the National Natural Science Foundation of China under Grant No. 62371038 and the Tianjin Natural Science Foundation under Grant No. 23JCQNJC00560.

\medskip
\section*{Conflict of Interest}
The authors declare no conflict of interest.

\medskip
\section*{Data Availability Statement}
The data that support the findings of this study are available from the corresponding author upon reasonable request.

\end{document}